\documentclass{aastex}
\usepackage{spr-astr-addons}
\usepackage{url}
\usepackage[bf, hang]{subfigure}

\RequirePackage{color}

\def\be{\begin{equation}}
\def\ee{\end{equation}}
\def\bea{\begin{eqnarray}}
\def\eea{\end{eqnarray}}
\def\beq{\begin{eqnarray}}
\def\eeq{\end{eqnarray}}

\begin{document}

\title{Gravitational and electromagnetic emission by magnetized coalescing binary systems}

\shorttitle{ Gravitational and electromagnetic emission by binary systems}
\shortauthors{S. Capozziello et al.}

\author{S. Capozziello\altaffilmark{1,2}, M. De
Laurentis\altaffilmark{1,2},} \and\author{ I. De
Martino\altaffilmark{3}, M. Formisano\altaffilmark{4}, D. Vernieri\altaffilmark{1}}


\altaffiltext{1}{Dipartimento di Scienze Fisiche, Università di
Napoli {}``Federico II''}
\altaffiltext{2}{INFN Sez. di Napoli, Compl. Univ. di
Monte S. Angelo, Edificio G, Via Cinthia, I-80126, Napoli, Italy. }
\altaffiltext{3}{Fisica Teorica, University of Salamanca, 37008 Salamanca, Spain}
\altaffiltext{4}{Dip. di Fisica, Universit\`a di Roma "La Sapienza", Piazzale Aldo
Moro 5, I-00185 Roma, Italy}

\begin{abstract}
We discuss the possibility to obtain an electromagnetic emission
accompanying the gravitational waves emitted in the coalescence of a compact
binary system. Motivated by the existence of black hole configurations with
open magnetic field lines along the rotation axis, we consider a magnetic
dipole in the system, the evolution of which leads to (i) electromagnetic
radiation, and (ii) a contribution to the gravitational radiation, the
luminosity of both being evaluated. Starting from the observations on
magnetars, we impose upper limits for both the electromagnetic emission and
the contribution of the magnetic dipole to the gravitational wave emission.
 Adopting this model for the evolution of neutron star binaries leading to
short gamma ray bursts, we compare the correction originated by the electromagnetic field to the gravitational waves emission,
 finding that they are comparable for particular values of the magnetic field and of the orbital radius of the binary system. 
 Finally we calculate the electromagnetic and gravitational wave energy outputs  which result comparable for some values of  
 magnetic field and  radius. 
\end{abstract}



\section{Introduction\label{uno}}

According to General Relativity, compact binary systems (composed of neutron
stars (NS), white dwarfs or black holes (BH)) are strong emitters of
gravitational waves (GWs), the ripples of space-time due to the presence of
accelerated masses (coalescence). As a consequence of the gradual inspiralling, the system
loses energy, linear and angular momentum (\cite{HulseTaylor,Nice:2002, Taylor, Stairs:2004, weisberg:2002}).
The coalescence of a compact binary system can be split into three distinct,
but not sharply delimited phases: the \emph{inspiral}, the \emph{
plunge/merger}and the \emph{ring-down}. For the
inspiral phase an accurate analytical description via the so-called
Post-Newtonian (PN) expansion has been developed (for a review see (\cite
{Blanchet}). Spin and mass quadrupolar contributions to this dynamics are
also well-known (\cite{BOC}), with inclusion of  leading order
spin-orbit (\cite{Ryan:1996,Majar:2008,Cornish:2010,Kidder:1993,Kidder:1995,O'Connell:2004,Rieth:1997,Will:2005,Zeng:2007}),
spin-spin (\cite{Wang:2007,Arun:2009,Klein:2010,Apostolatos:1995,Apostolatos:1996,Majar:2009})
and mass quadrupole mass monopole (\cite{Poisson:1998, Racine:2008}) couplings.

Similarly to the inspiral, the ring-down phase admits a
perturbative analytical model which describes the damped oscillations of the
single compact object resulting from the binary coalescence (\cite{ringdown}).
The plunge/merger phase is however non-perturbative, having no analytical
description for generic systems, therefore it is studied through numerical
simulations.

Templates on binary inspiral  (\cite{Buonanno:1999, Buonanno:2000,
Buonanno:2003, Buonanno:2006, Cutler:1993, Damour:1998,
Damour:2000, Damour:2001,Damour:2002, Damour:2003, Finn:1993, Flanagan:1998, Pan:2004}) and robust search algorithms were developed for GWs sources observable with
ground-based (\cite{LIGO,LIGO2,alggen1,alggen2}) and space-based interferometers (\cite{LISA, LISA2}).

In addition to GWs emission, these systems emit electromagnetic radiation(\cite{noi1,noi2}). This emission is due to dynamics of system and to behavior of plasma around the system. There are many different systems that emit gravitational and electromagnetic waves, and in many case the electromagnetic emission is very energetic (AGNs, QSOs, GRBs).

Pulsars with surface magnetic fields in
the range $10^{8\div 12}$ G and magnetars with surface magnetic fields up to
$10^{14\div 15}$ G have been reported (\cite{Duncan:1992, Sinha, Kouveliotou:1998, Thompson:1995, Usov:1992, Vasisht:1997, Woods:1999}), as summarized in (\cite
{Sinha}). An upper limit of $10^{18}$ G was also advanced from the
requirement to avoid exotic structure in the neutron stars (\cite{Sinha}). An
interaction of the magnetic dipoles modifies the gravitational radiation
leaving the system, and hence the back-reaction to the orbit is slightly
different. Limits on magnetar magnetic fields render these corrections to
the second or even higher PN order.
Neutron star binary coalescence, leading to a final black hole with a torus
is a configuration believed to be the progenitor of short GRBs. In the black
hole-torus central engine part of the torus matter can be converted into
radiation, leading to the observed bursts (\cite{Belczynski:2008}). Fully
general-relativistic simulations of binary neutron star coalescence leading
to this configuration have been presented in (\cite{Rezzolla}), with a
detailed and accurate study of the torus. It was found that unequal mass
neutron star binaries led to a larger torus and an empirical formula in
terms of the mass ratio and total mass was advanced for the torus mass.
Nevertheless, the effects of the magnetic fields were not included in the
treatment of Ref. (\cite{Rezzolla}).
Many efforts are done in order to study the electromagnetic emission from these systems, in particular 
from a computational point of view. It was showed how the dynamics of the binary systems
 induce the electromagnetic emission and its link with GWs emission (\cite{Pal2009,Pal2010,And2008}).

In this paper we investigate a simple model for the inspiral, which also
takes into account the magnetic field. Rather then monitoring the
complicated evolution of the individual magnetic momenta of a binary neutron
star system, we will introduce a global measure of the electromagnetic
properties of the system, given by a single magnetic dipole.

Gravitational radiation transports away energy, linear and orbital angular
momentum from the system. While the total angular momentum (defined as the
sum of the orbital angular momentum and possible individual spins of the
NS) exhibits instantaneous changes during orbital evolution, its
secular evolution over timescales larger than the orbital period is a pure
decrease in magnitude, with the direction kept fixed (\cite{ACST}).

Having in mind (i) the configuration of open field lines aligned with the
spin of the final compact object resulting from the coalescence, and (ii)
the property of the gravitational radiation to conserve the direction of the
total angular momentum on timescales longer than the orbital period, we
chose the magnetic field aligned with the total angular momentum of the
system.

In Sec. 2, we present an overview on GWs emission from coalescing binary
systems. In Sec. 3, we introduce and discuss the magnetic field and we also
calculate the correction induced in the GW luminosity and amplitude. For
comparison, in Sec. 4, we present the luminosity of the accompanying
electromagnetic radiation. These are compared and the conclusions presented
in Sec. 5.

\section{Gravitational wave emission from coalescing binaries\label{due}}

Let us summarize some  basic concepts s of the gravitational
radiation generated by the evolution of a  compact binary system. A wide discussion about
the GWs emission can be found in literature (\cite{Maggiore, Misner, wein}).
The starting point for any discussion of GWs is the approximation to General
Relativity (GR), when the space-time metric can be treated as deviating only
slightly from a flat metric (\cite{Maggiore, Shapiro,Misner}). We start from linearized Einstein equation
\begin{equation}
\square \overline{h}_{\mu \nu }=-\frac{16\pi G}{c^{4}}\,T_{\mu \nu }~,
\end{equation}%
where $\Box =\eta _{\mu \nu }\partial ^{\mu }\partial ^{\nu }$ is the
d'Alembertian, expressed in terms of the trace reversed metric
perturbation
\begin{equation}
\overline{h}^{\mu \nu }={h}^{\mu \nu }-\frac{1}{2}\eta ^{\mu \nu }h~,
\end{equation}%
where $h_{\mu \nu }=g_{\mu \nu }-\eta _{\mu \nu }$ and $h=\eta _{\mu \nu
}h^{\mu \nu }$ (such that $\overline{h}=-h$). Due to the imposed Lorentz
(harmonic) gauge $\partial _{\nu }\overline{h}^{\mu \nu }=0$, only 6 of the
10 components of $h_{\mu \nu }$ are independent. By setting $T_{\mu \nu }=0$
we obtain the vacuum wave equation:
\begin{equation}
\square \overline{h}_{\mu \nu }=0\,.  \label{eq:20}
\end{equation}%
GWs thus propagate at the speed of light.

We evaluate the leading order contribution to the metric perturbations under
the assumption that the internal motions of the source are slow compared to
the speed of light. We also assume that the source's self-gravity is
negligible. Let us first introduce the momenta of the mass density
\begin{eqnarray}
M &=&\frac{1}{c^{2}}\int d^{3}x\,T^{00}(t,\mathbf{x})\,,  \label{eq:99} \\
M^{i} &=&\frac{1}{c^{2}}\int d^{3}x\,T^{00}(t,\mathbf{x})\,x^{i}\,,
\label{eq:100} \\
M^{ij} &=&\frac{1}{c^{2}}\int d^{3}x\,T^{00}(t,\mathbf{x})\,x^{i}\,x^{j}~,
\label{eq:59}
\end{eqnarray}%
and impose the conservation law $\partial _{\mu }T^{\mu \nu }=0$, valid in
linearized gravity. For sources having strong gravitational field, as NSs or
BHs, the mass density depends also on the binding energy, however for weak
fields and small velocities $T^{00}/c^{2}$ reduces to the rest-mass density $%
\rho $. Denoting the field $h_{\mu \nu }$ which satisfies the transverse and
traceless gauge conditions (\cite{wein}), as transverse-traceless tensor $%
h_{ij}^{TT}$, it is convenient to compute from it explicitly the two
indipendent polarization states $h_{+}$ and $h_{\times }$ (\cite{Buonanno}).

In terms of the traceless quadrupole tensor
\begin{equation}
Q_{ij}=M_{ij}-\frac{1}{3}\,\delta _{ij}\,\mathcal{M},  \label{eq:quadrupole}
\end{equation}%
($\mathcal{M}$ denoting the trace of $M_{ij}$), the leading order expression
of the total radiated power is
\begin{equation}
P=\frac{G}{5c^{5}}\,\langle \dddot{Q}_{ij}\,\dddot{Q}_{ij}\rangle \,.
\label{eq:73}
\end{equation}
Henceforth, we will apply this formalism for a compact binary system with
masses $m_{1}$ and $m_{2}$, total mass $M=m_{1}+m_{2}$ and reduced mass $\mu
=m_{1}\,m_{2}/(m_{1}+m_{2})$.
With the plane of motion chosen as the $x$-$y$ plane and in the circular
motion approximation the reduced mass particle has the coordinates $\mathbf{r%
}=\left( x,y,z\right) $ given by
\begin{equation}
x(t)=r\,\cos \omega \,t\,,\quad y(t)=r\,\sin \omega \,t\,,\quad z(t)=0\,,
\end{equation}%
$r$ being the relative distance between the two bodies. With $\rho =\mu
\delta (\mathbf{r})$ we find $M^{ij}=\mu x^{i}\,x^{j}$ with the nontrivial
components
\begin{eqnarray}
M_{11} &=&\frac{1}{2}\,\mu \,r^{2}\,(1+\cos 2\omega t)\,,  \notag \\
M_{22} &=&\frac{1}{2}\,\mu \,r^{2}\,(1-\cos 2\omega t)\,,  \notag \\
M_{12} &=&\frac{1}{2}\,\mu \,r^{2}\,\sin 2\omega t\,.  \label{eq:80}
\end{eqnarray}%
From which we obtain
\begin{equation}
h_{+}(t)=\frac{4G}{c^{4}d}\,\mu \,r^{2}\,\omega ^{2}\,\frac{(1+\cos
^{2}\theta )}{2}\,\cos (2\omega \,t)\,\,,
\end{equation}%
\begin{equation}
h_{\times }(t)=\frac{4G}{c^{4}d}\,\mu \,r^{2}\,\omega ^{2}\,\cos \theta
\,\sin (2\omega \,t)\,.  \label{eq:81}
\end{equation}%
Here the time is shifted in order to eliminate the dependence on $\phi $
(\cite{Buonanno}). From Eq. (\ref{eq:73}) the total radiated power emerges as,
\begin{equation}
P=\frac{32}{5}\,\frac{G\,\mu ^{2}\,r^{4}\,\omega ^{6}}{c^{5}}\,.
\label{eq:83}
\end{equation}

In order to compensate for the loss of energy by GW emission, the radial
separation $r$ between the two bodies must decrease. The orbital frequency
and consequently the GW frequency also changes in time, and can be derived
using from the balance equation
\begin{equation}
\frac{dE_{\mathrm{orbit}}}{dt}=-P\,.  \label{eq:84}
\end{equation}

\section{Gravitomagnetic corrections on gravitational wave emission\label%
{tre}}

In this section we introduce the magnetic field with the desired structure,
namely aligned with the total angular momentum $\mathbf{J}$\ of the system.

\subsection{Gravitomagnetic-electromagnetic analogy}

The \emph{gravitomagnetic} potential characterizing the compact binary is $%
{{A_{\mu }}=}\left( 0,\mathbf{A}\right) $, with
\begin{equation}
\mathbf{A}={{\alpha \frac{\mathbf{J}\times \mathbf{r}}{{{r^{3}}}}}~},
\label{potVett}
\end{equation}%
where $\alpha =G/c$ \cite{tartaglia}, and $\mathbf{r}$ is the separation between the bodies of the binary system.
Due to the fact that the motion of the binary system is entirely confined into the orbital plane, we evaluate all the physical quantities
of interest in the plane itself. This procedure is analogue to that used in GWs emission calculations, where one reduces the problem only to the planar relative motion.
The gravitomagnetic field is found as
\begin{equation}
\mathbf{B}=\mathbf{{\nabla }\times \mathbf{A}}=\frac{2\alpha }{r^{3}}\mathbf{%
J~}.  \label{B}
\end{equation}%
We have used the assumption, that neutron stars rotate slowly, therefore the
proper spin contributions to $\mathbf{J}$\ are negligible, also that the
relativistic (PN and 2PN corrections) are aligned to the Newtonian orbital
angular momentum, thus $\mathbf{J\cdot r}=0$ (for a motion confined to the $x
$-$y$ plane the only nonvanishing component of $\mathbf{J}$ is $J_{z}\equiv J
$).
Similarly, a \emph{gravitoelectric} field emerges as
\begin{equation}
\mathbf{E}=\frac{\alpha }{{{r^{3}}}}\mathbf{v}\times \mathbf{J~}.  \label{E}
\end{equation}%
and they combine to a gravito-electromagnetic tensor ${{F_{\mu \nu }}={%
\partial _{\mu }}{A_{\nu }}-{\partial _{\nu }}{A_{\mu }}}$.

The remarkable property of this gravitomagnetic field $\mathbf{B}$ is its
alignement to $\mathbf{J}$, a property we seek for the \emph{magnetic}
field in our model. Therefore we will introduce an \emph{electromagnetic}
field analogous to the gravitoelectric and gravitomagnetic fields, by
changing $\alpha =G/c$ into $\alpha =\mu_0/4\pi$,
where $\mu _{0}$ is the vacuum magnetic permeability. In what follows, by $%
\alpha $ we mean this expression and we refer to $\mathbf{A},\mathbf{B},%
\mathbf{E},F_{\mu \nu }$ as true electromagnetic quantities.

\subsection{Gravitational wave energy loss and polarizations}

We characterize the source by the total stress-energy tensor%
\begin{equation}
{T^{\mu \nu }}=T_{PF}^{\mu \nu }+T_{EM}^{\mu \nu }\ .  \label{EnergyTensor}
\end{equation}%
composed by a perfect fluid part
\begin{equation}
T_{PF}^{\mu \nu }=\left( {\rho +\frac{p}{{{c^{2}}}}}\right) {u^{\mu }}{%
u^{\nu }}-p{g^{\mu \nu }}
\end{equation}%
and an electromagnetic part%
\begin{equation}
T_{EM}^{\mu \nu }=\frac{1}{\mu _{0}}\left( {{F^{\mu \alpha }}{F^{\nu \beta }}%
{\eta _{\alpha \beta }}-\frac{1}{4}{\eta ^{\mu \nu }}{F_{\sigma \lambda }}{%
F^{\sigma \lambda }}}\right) ~.
\end{equation}%
In order to calculate the momenta (\ref{eq:99})-(\ref{eq:59}) we need the
00-components of these tensors. The 00-component of the perfect fluid
contribution in a comoving frame defined as
\begin{equation}
u^{\alpha }=\left( {{u^{0}},0,0,0}\right) ~,\quad {u_{0}}{u^{0}}={c^{2}}~
\end{equation}%
simply becomes
\begin{equation}
{T_{PF}^{00}=T_{00}^{PF}=\rho {c^{2}}},  \label{EnergyTensorGrav}
\end{equation}%
while for the electromagnetic contribution in the system in which the
magnetic field is along the $z$-axis [such that $\mathbf{J}=(0,0,{J})$]$~$we
find%
\begin{eqnarray}
T_{EM}^{00} &=&-\mu^{2}\alpha^3\epsilon _{EM}~,
\label{EnergyTensorEl} \\
\epsilon _{EM} &=& \left( \frac{{J}}{
{\mu r^3c}}\right) ^{2}\left[ 2c^2-{\frac{3}{2}} r^2\omega^2\right] ~.  \label{EMcorr}
\end{eqnarray}%
Despite apparences, $%
T_{EM}^{00}$ does not depend on masses, but it was written this way for an
easy PN order evaluation: $\mathcal{O}\left( J/\mu rc\right) ^{2}=\mathcal{O}%
\left( v/c\right) ^{2}=\mathcal{O}\left( GM/c^{2}r\right)$. We also remark, that $%
\mathcal{O}\left( r\omega /c\right) ^{2}=\mathcal{O}\left( v/c\right) ^{2}$,
therefore the respective term of the expression (\ref{EMcorr}) could be safely
dropped.

Inserting
\begin{equation}
{{T^{00}}=T_{PF}^{00}+T_{EM}^{00}=\rho {c^{2}}}-\mu^{2}\alpha^3\epsilon _{EM}~  \label{E-tensor}
\end{equation}%
into (\ref{eq:59}),  the  corrections to the gravitational waves emission can be derived.  In  a binary system,   the presence of an additional dipole-electromagnetic field has to be 
taken into account. The correction can be derived by considering also the electromagnetic stress-energy tensor  beside   
the perfect fluid one. stress energy tensor. Thanks to the gravito-magnetic 
ormalism discussed above, we can consider the electromagnetic contribution under the same standard  of the  gravitational one: this means  to take into account  the 00-component of the total stress-energy tensor given by the  gravitational part (the mass density) and the electromagnetic contribution. In order to get the momenta 
(obtained by integrating the total stress-energy tensor in the volume where  the binary system is contained) 
we have to compute an integral consisting of  a gravitational and  an   electromagnetic part:
\begin{eqnarray}
&&{M_{ij}} = \underbrace {\int {{d^3}x} \rho \left( x \right){x_i}{x_j}}_{{\rm{Gravitational \rm{ \ }part}}} + \nonumber\\ &&
{\rm{    \quad \quad  \quad      }} - \underbrace {{\mu ^2}{\alpha ^3}{{\left( {\frac{{J}}{{\mu {c^2}}}} \right)}^2}\int {{d^3}x\frac{1}{{{r^6}}}\left( {\frac{3}{2}{r^2}{\omega ^2} - 2c^2} \right){x_i}{x_j}} }_{{\rm{Electromagnetic\rm{\ } part}}}\nonumber\\
\end{eqnarray}
where $\mu$ is the reduced mass of the binary system. We obtain
\begin{equation}
{M_{ij}}=\frac{\mathcal{M}}{2}\left( {%
\begin{array}{ccc}
{1+\cos \left( {2\omega t}\right) } & {\sin \left( {2\omega t}\right) } & 0
\\
{\sin \left( {2\omega t}\right) } & {1-\cos \left( {2\omega t}\right) } & 0
\\
0 & 0 & 0%
\end{array}%
}\right) ~,  \label{momemnta}
\end{equation}%
where $\mathcal{M}$ is its trace, given by%
\begin{equation}
\mathcal{M} = \mu {r^2}\left( {1 - \frac{{4\pi \mu {\alpha ^3}}}{{3{r^3}}}{{\left( {\frac{J}{{\mu c^2}}} \right)}^2}\left[ {2{c^2} - \frac{3}{2}{r^2}{\omega ^2}} \right]} \right).\label{MTensorTrace}
\end{equation}%
The nontrivial quadrupole tensor components are
\begin{eqnarray}
{Q_{11}} &=&\frac{1}{6}\mathcal{M}\left[ 1{+3\cos \left( {2\omega t}\right) }%
\right] ~,  \notag \\
{Q_{22}} &=&\frac{1}{6}\mathcal{M}\left[ 1{-3\cos \left( {2\omega t}\right) }%
\right] ~,  \notag \\
{Q_{33}} &=&-\frac{1}{3}\mathcal{M}~,  \notag \\
{Q_{12}} &=&\frac{1}{2}\mathcal{M}\sin \left( {2\omega t}\right) ~.
\end{eqnarray}%
Inserting them into Eq. (\ref{eq:84}) we obtain the loss of energy
\begin{eqnarray}
{\left( \frac{dE}{dt} \right)_{GW_T}} &=& \left( {\frac{{dE}}{{dt}}} \right)_{GW}+\left( {\frac{{dE}}{{dt}}} \right)_{GW_B}  =  \nonumber \\\nonumber \\
&=&- \frac{{32G{\mu ^2}{r^4}{\omega ^6}}}{{5{c^5}}}\left( {1 - \frac{{8\pi {r^3}}}{{15{c^2}}}\mu {\alpha ^4}{\epsilon _{EM}}} \right)\,,\nonumber\\
\end{eqnarray}%
where $\displaystyle{\left( {\frac{{dE}}{{dt}}} \right)_{GW_T}}$ stands to indicate the total energy output of gravitational waves emission due to the pure gravitational term $\displaystyle{\left( {\frac{{dE}}{{dt}}} \right)_{GW}}$, and the electromagnetic correction $\displaystyle{\left( {\frac{{dE}}{{dt}}} \right)_{GW_B}}$.
The polarizations result to be
\begin{eqnarray}
{h_ + }& =& \frac{{G\mu{r^2}{\omega ^2}}}{{{c^4}d}}\left( {1 - \frac{{4\pi \mu {r^3}{\alpha ^4}}}{{3c}}{\epsilon_{EM}}} \right)\times \nonumber\\ &&
\times \left[ {3 + \cos (2\theta )} \right]\cos (2\omega t),  \label{h+}
\end{eqnarray}%

\begin{equation}
{h_ \times } =  - \frac{G\mu {r^2}{\omega ^2}}{{c^4}d}\left( {1 - \frac{{4\pi \mu {r^3}{\alpha ^4}}}{{3c}}{\epsilon_{EM}}} \right)\cos \theta \sin (2\omega t).  \label{hx}
\end{equation}

\begin{figure*}[!h]
\caption{Profiles of the energy loss in  time when we increase the orbital radius ($r$) and the total emitted gravitational energy with respect to the electromagnetic correction to GWs emission: (a)  the magnetic field is $B=10^{12}$G; (b)   $B=10^{15}$G; (c)  $B=10^{18}$G.}\label{fig:energy}
\subfigure[]{\includegraphics[scale=0.41]{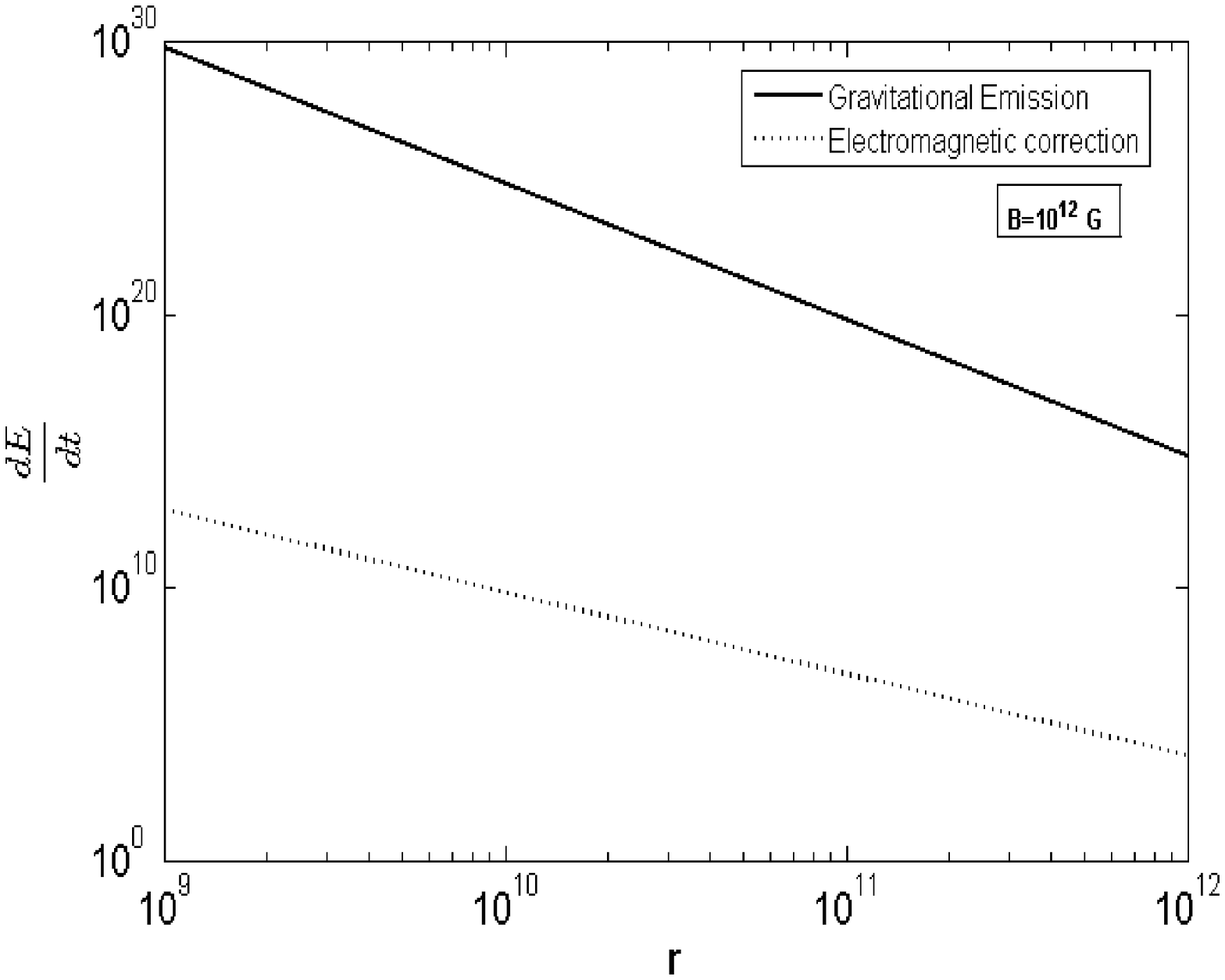}}
\subfigure[]{\includegraphics[scale=0.41]{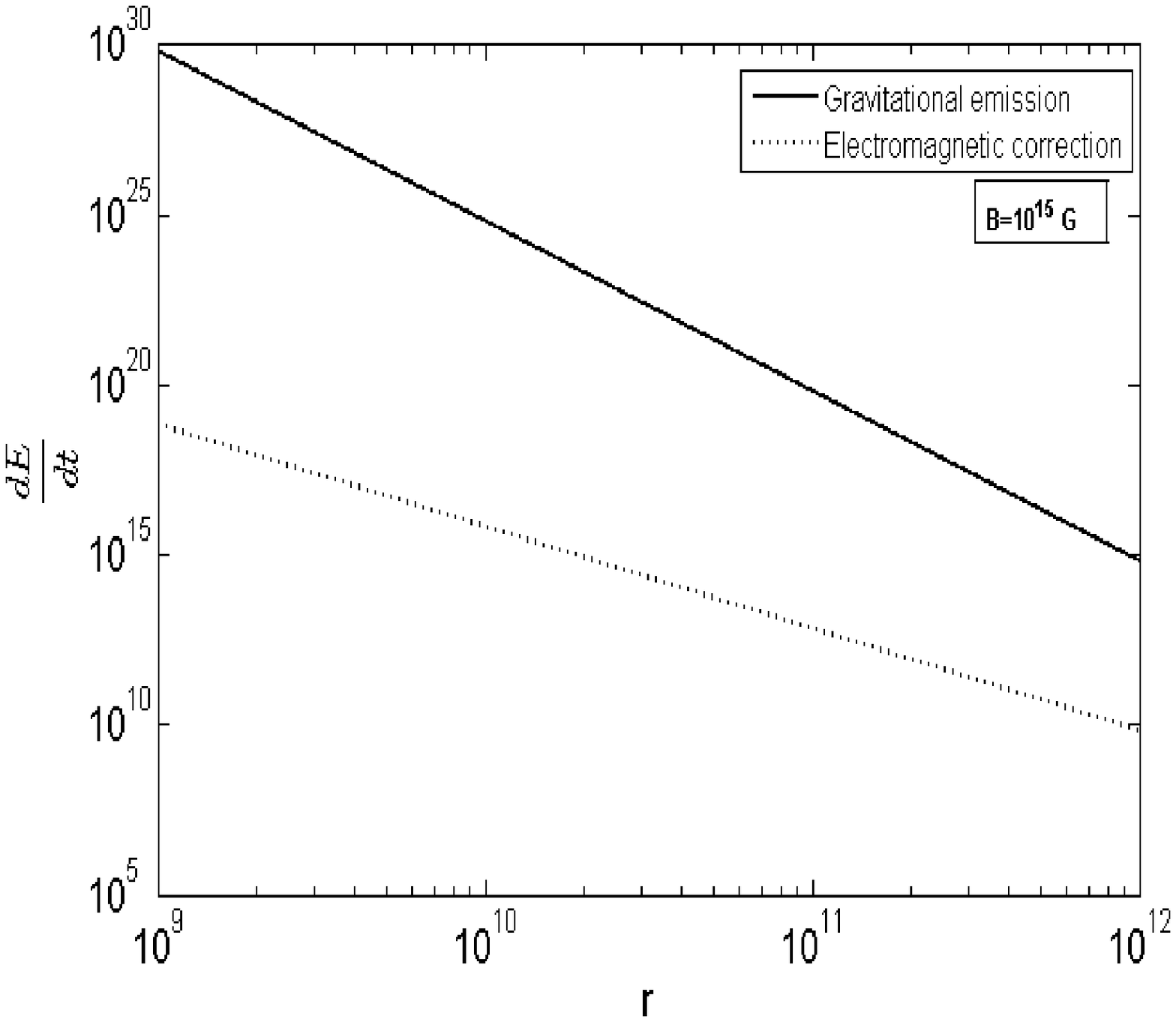}} \\
\subfigure[]{\includegraphics[scale=0.41]{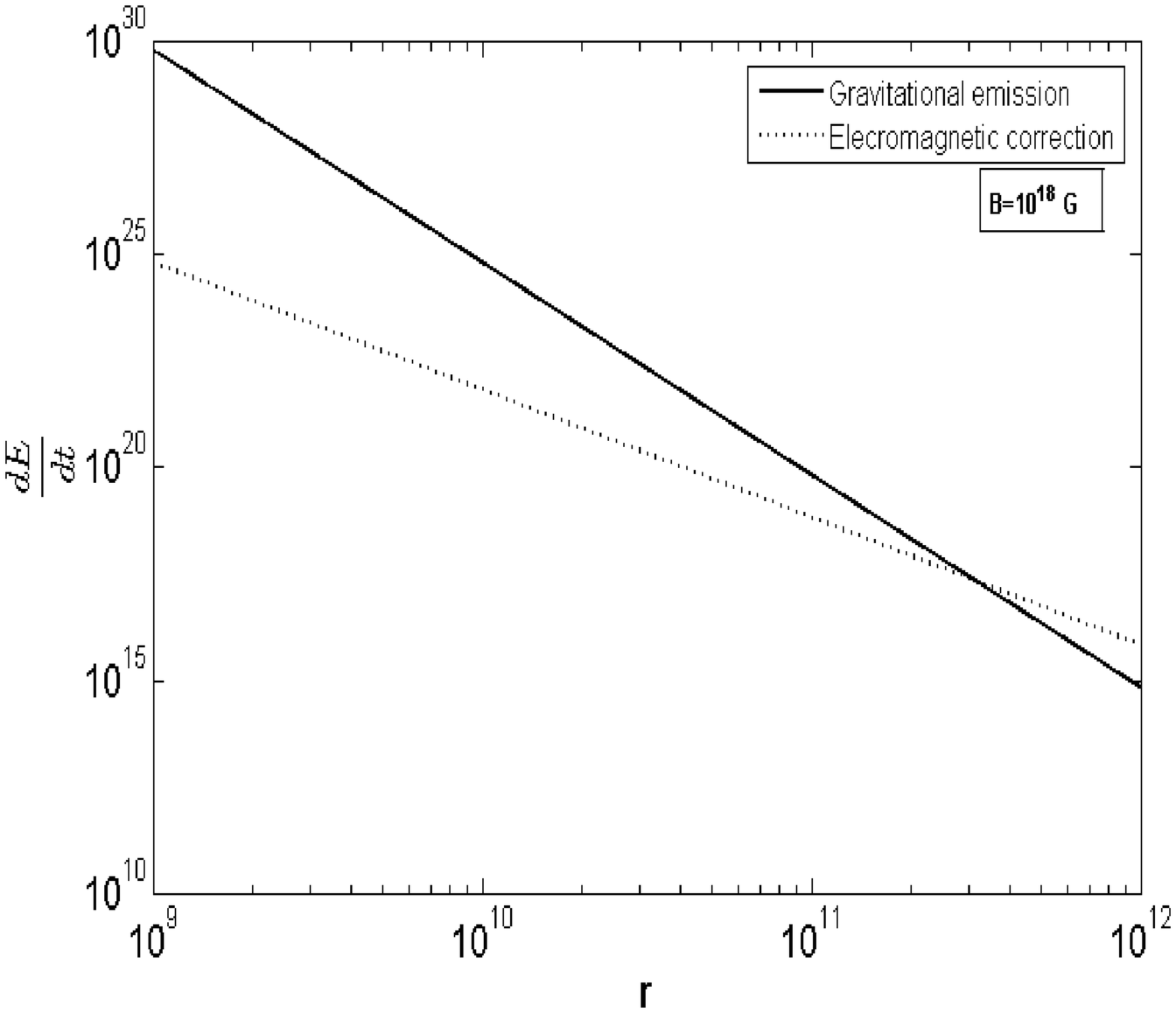}}
\end{figure*}

Calculating $h_+$ and $h_\times$ for a NS binary system using
\begin{equation*}
\begin{array}{l}
 r = {10^{9}\div10^{12}}{\rm{m}},{\rm{ \ \    }}d \sim {10^{19}}{\rm{m}}{\rm{,   \ \   }}{m_1} = {m_2} = 1,4{M_ \odot },
 \end{array}
\end{equation*}

after a few calculations we obtain that the corrective electromagnetic term is initially negligible with respect to the gravitational one, but it becomes not negligible when we increase the orbital radius, as well as the magnetic field, as shown in Fig \ref{fig:energy}.
Similar results can be obtained also for an equal-mass BH or White Dwarfs binary system for which the corrective terms are negligible.
However, it is possible to obtain for a binary system with a very strong dipole magnetic momentum, an electromagnetic contribution to GWs emission which is comparable to the standard gravitational one. \\


\section{Electromagnetic emission by the magnetized binary}\label{quattro}


\begin{table*}[t]
\centering
\caption{The columns of the table, from left to right, indicate the radial
separation between the two bodies ($r$), the orbital frequency ($\protect%
\omega$), the electromagnetic energy emitted in the unit-time $\left( \dfrac{%
dE_{EM}}{dt}\right)$, the loss of energy in gravitational emission $\left(%
\dfrac{dE_{GW}}{dt} \right)$ and its electromagnetic correction $\left( \dfrac{dE_{GW_B}%
}{dt}\right) $. The ranges are for magnetic fields $B=(10^{12}\div10^{18})$
G. }\label{tab:results}
\begin{tabular}{@{}ccccc@{}}
\tableline
\\
$r$ (m) & $\omega $ (Hz) & $-\left( \frac{dE}{dt}\right) _{EM}$ (erg/s) & $%
-\left( \frac{dE}{dt}\right) _{GW}$ (erg/s) & $-\left( \frac{dE}{dt}\right)
_{GW_B}$ (erg/s) \\ \tableline
$10^{9}$ & $3.04\times 10^{-4}$ & $1.14\times 10^{21}\div 1.14\times 10^{33}$
& $6.72\times 10^{29}$ & $6.88\times 10^{12}\div 6.88\times 10^{24}$ \\
$10^{10}$ & $9.61\times 10^{-6}$ & $3.60\times 10^{20}\div 3.60\times
10^{32} $ & $6.72\times 10^{24}$ & $6.88\times 10^{9}\div 6.88\times
10^{21}$ \\
$10^{11}$ & $3.04\times 10^{-7}$ & $1.14\times 10^{20}\div 1.14\times
10^{32} $ & $6.72\times 10^{19}$ & $6.88\times 10^{6}\div 6.88\times
10^{18}$ \\
$10^{12}$ & $9.61\times 10^{-9}$ & $3.60\times 10^{19}\div 3.60\times
10^{31} $ & $6.72\times 10^{14}$ & $6.88\times 10^{3}\div 1.64\times
10^{15}$\\
\tableline
\end{tabular}
\end{table*}

Now we calculate the electromagnetic energy emission by binary systems in
order to compare it to GWs energy emission. We consider the electromagnetic
tensor components in order to calculate the time energy emission:
\begin{equation}  \label{TE01}
{T_E^{01} = -2\alpha \dot x\left( t \right){{\left( {\frac{{\alpha{J}}}{{{c%
}{r^3}}}} \right)}^2}},
\end{equation}
\begin{equation}  \label{TE02}
{T_E^{02} = -2\alpha \dot y\left( t \right){{\left( {\frac{{\alpha{J}}}{{{c%
}{r^3}}}} \right)}^2}},
\end{equation}
and finally
\begin{equation}  \label{TE03}
{T_E^{03} = 0}.
\end{equation}
Calculating the Poynting vector $\vec P$, whose components are $%
P^i=cT^{0i}_E $, it is possible to obtain the electromagnetic energy emitted
in the unit-time as

\begin{equation*}
\frac{{d{E_{EM}}}}{{dt}}=-\int_{\Sigma _{\mu }}\vec{P}\cdot d\vec{S},
\end{equation*}%
where $\Sigma _{\mu }$ is the surface of the reduced mass object, which
gives
\begin{equation}
\frac{{d{E_{EM}}}}{{dt}}=-{2\alpha c{r}\omega }{\left( {\frac{{\alpha {J}}}{{%
{c}{r^{3}}}}}\right) ^{2}}\Sigma _{\mu }.  \label{EnergyEM}
\end{equation}%
\newline
From eq. \eqref{B}, it is possible to get the dipole magnetic momentum $%
J $:
\begin{equation}
{J}=\frac{{{B}{r^{3}}}}{{2\alpha }}.  \label{BSz}
\end{equation}%
Now we are able to compare the different contributions to energy emission:
the standard GWs term, the electromagnetic correction to it and then the
pure electromagnetic one. The results are shown in Tab. \ref{tab:results}.

\section{Conclusions}
\label{cinque}

GWs science has entered a new era and the recent years have been characterized
by several major advances. For what concerns the most promising GW sources
for ground-based and space-based detectors, notably, binary systems composed
of NS and BHs, our understanding of the two-body problem and the GW
generation problem has improved significantly. The best-developed analytic
approximation  in General Relativity is undoubtable the post-Newtonian
method. In this paper, we have calculated the electromagnetic corrections to
the GWs emitted by a coalescing binary system. We have considered an
electromagnetic  dipole-type field and calculated the electromagnetic
contribution to the stress-energy tensor. We have obtained a correction to
the standard gravitational energy-loss which becomes null setting the
magnetic field to zero.  In general, the  electromagnetic correction term  is negligible with respect to the gravitational one.
This result holds also for  black hole and  white dwarfs binary systems.  However, there could be  
 coalescing binary systems with a large dipole magnetic momentum  and a  large orbital radius where the
 electromagnetic correction to GWs emission is relevant. In particular, the electromagnetic energy emitted by these binary systems is comparable to  the gravitational wave output for given values of the magnetic field and of the orbital radius. Revealing such a phenomenon by observations could be an electromagnetic  signature for the  gravitational waves emission.

\bibliographystyle{spr-mp-nameyear-cnd}

\end{document}